\documentclass[twocolumn,english,aps,prl,showpacs,superscriptaddress]{revtex4-2}
\usepackage[T1]{fontenc}
\usepackage[latin9]{inputenc}
\usepackage{geometry}
\usepackage[x11names]{xcolor}
\usepackage{babel}
\usepackage{units}
\usepackage{amsmath}
\usepackage{amssymb}
\usepackage{mathtools} 
\usepackage{braket} 
\usepackage{stmaryrd}
\usepackage{graphicx}
\usepackage{wasysym}
\usepackage{hyperref}
\usepackage{orcidlink}
\makeatletter

\newcommand{\BL}[1]{{\color{black}#1}}

\IfFileExists{lmodern.sty}{\usepackage{lmodern}}{}
\usepackage{babel}

\def\equationautorefname~#1\null{Equation (#1)\null}

\makeatother

\begin{document}
\title{Suppressing the decoherence of alkali-metal spins at low magnetic fields}
\date{\today}

\author{Mark Dikopoltsev\orcidlink{0000-0001-9842-2575}}
\email{mark.dikopoltsev@mail.huji.ac.il}
\address{Institute of Applied Physics, The Faculty of Science, The Center for Nanoscience and Nanotechnology, The Hebrew University of Jerusalem, Jerusalem 9190401, Israel.}
\address{Rafael Ltd, 31021, Haifa, Israel.}
\author{Avraham Berrebi\orcidlink{0009-0001-4074-4413}}
\address{Institute of Applied Physics, The Faculty of Science, The Center for Nanoscience and Nanotechnology, The Hebrew University of Jerusalem, Jerusalem 9190401, Israel.}
\author{Uriel Levy\orcidlink{0000-0002-5918-1876}}
\address{Institute of Applied Physics, The Faculty of Science, The Center for Nanoscience and Nanotechnology, The Hebrew University of Jerusalem, Jerusalem 9190401, Israel.}
\author{Or Katz\orcidlink{0000-0001-7634-1993}}
\email{or.katz@cornell.edu}
\address{School of Applied and Engineering Physics, Cornell University, Ithaca, NY 14853.}

\begin{abstract}
Interactions of electron spins with rotational degrees of freedom during collisions or with external fields are fundamental processes that limit the coherence time of spin gases. We experimentally study the decoherence of hot cesium spins dominated by spin rotation-interaction during binary collisions with N$_2$ molecules or by absorption of near-resonant light. We report an order of magnitude suppression of the spin decoherence rate by either of those processes at low magnetic fields. This work extends the use of magnetic fields as a control knob, not only to suppress decoherence from random spin-conserving processes in the Spin-Exchange Relaxation Free (SERF) regime but also to suppress processes that relax electron spins rather than conserve them.

 \end{abstract}
\maketitle
Ensembles of alkali-metal spins are prominent physical systems. They feature strong coupling to optical and magnetic fields and can be isolated from the environment for considerably long times at or above room-temperature. Therefore, they have wide-spread and evolving applications in fields such as quantum sensing and precision magnetometry  \cite{budker2007optical,kominis2003subfemtotesla,dang2010ultrahigh,balabas2006magnetometry,walker2016spin,ledbetter2008spin,xu2006magnetic,theis2011parahydrogen}, in searches of new physics 
\cite{jiang2021search,bloch2020axion,bloch2022new,vasilakis2009limits,budker2014proposal,safronova2018search,budker2022quantum,afach2021search,bloch2023constraints}, in interfacing and polarizing spins of noble-gases for imaging and fundamental studies \cite{walker1997spin,wagshul1994laser,gentile2017optically,shaham2022strong,katz2021coupling,chupp2001medical,coulter1988neutron}, in coupling to opto-mechanical systems \cite{moller2017quantum,thomas2021entanglement,khalili2018overcoming}, in quantum information applications \cite{hammerer2010quantum,bao2020spin,katz2020long,kong2020measurement,katz2018light,hosseini2011high,guarrera2021spin,mouloudakis2022effects,buser2022single,hammerer2010quantum,julsgaard2001experimental} and recently also in studies of new phases of matter \cite{horowicz2021critical}. 

The coupling of the electron spin to external fields also sets the limit for which the spin state can be isolated from the environment before relaxing, affecting the performance of the aforementioned applications. The prominent relaxation mechanisms of alkali-metal spins originate from processes which couple predominantly to the valence electron spin, during collisions, by the action of external fields, or at the walls of the enclosure that holds the gas \cite{appelt1998theory}. The latter can be suppressed by introducing buffer gas, typically consisting of mixtures of noble-gas atoms or diatomic molecules. This gas renders the motion diffusive (rather than ballistic) and slows-down the collision rate of alkali-metal atoms with the glass walls \cite{happer2010optically}. However, buffer gas also acts to relax the alkali-metal's spin by the spin-rotation interaction, which couples the electron spin to the rotational angular momentum during collisions \cite{wu1985spin,walker1989estimates,walker1997spin2,happer2010optically,wagshul1994laser}. Other relaxation mechanisms are associated with the action of external fields. For instance, optical fields tuned near the atomic transitions which are used, e.g.~to dispersively probe the spin state of alkali atoms or confine their motion in optical traps, can be absorbed by the atoms and alter or relax their spin state
\cite{hammerer2004light,hammerer2010quantum,kong2020measurement}.

The rates at which different relaxation mechanisms practically affect alkali-metal spins depend on several parameters, including the form of the perturbation, the spin quantum numbers, and external fields such as the magnetic field and its orientation. The latter, determines the quantization axis. Early works have theoretically derived relaxation rates for different spin multipoles under general forms of perturbations, with examples including magnetic-field and electric-field gradient noise \cite{happer1970multipole,happer1972optical,bouchiat1963relaxation,happer1967effective}. Regarding the magnetic dipole moment of spins \cite{happer1970multipole}, spins oriented along the quantization axis exhibit a characteristic "longitudinal" spin lifetime $T_1$ \cite{appelt1998theory,kadlecek2001spin,graf2005relaxation}. Conversely, spins oriented transverse to the quantization axis have a coherence time $T_2$, typically much shorter than the spin lifetime, especially for dense alkali-metal vapor. The reduced coherence time partially originates from processes such as random spin-exchange collisions between pairs of alkali-metal atoms, which predominantly decohere the spins but weakly affect their lifetime \cite{happer1977effect,appelt1998theory,happer2010optically}. 
Nevertheless, decoherence due to random \textit{spin-conserving} collisions can be usefully suppressed via control over the magnetic field magnitude in the $\mu$G-G range; It is utilized to suppress the decoherence caused by the random spin-exchange collisions and to drive the vapor into the Spin-Exchange Relaxation Free (SERF) regime \cite{happer1977effect,happer1973spin,savukov2005effects,korver2013suppression,katz2013nonlinear,chalupczak2014spin,gartman2018linear}. Yet, the direct effect of the magnetic field magnitude on decoherence by processes which do not conserve the spins, such as spin-rotation interaction in collisions with buffer gas or absorption of light, has not been investigated experimentally below hundreds of Gauss \cite{walter2002magnetic,kadlecek1998field,erickson2000spin}.

Here we study decoherence that originates from processes that do not conserve the spin, and its dependence on the applied magnetic field. We report an order of magnitude suppression of the decoherence rate by the spin-rotation interaction during collisions of cesium vapor with $\textrm{N}_2$ molecules, and further show that the suppression is robust for a wide range of densities and degrees of spin-polarization. We additionally study the decoherence rate by the absorption of near-resonant light and find an order of magnitude suppression at low magnetic fields. We present a simple model that aligns well with our measurements and highlight applications where magnetic fields can serve as an effective tool to suppress relaxation.

\begin{figure}[t]
\begin{centering}
\includegraphics[width=8.6cm]{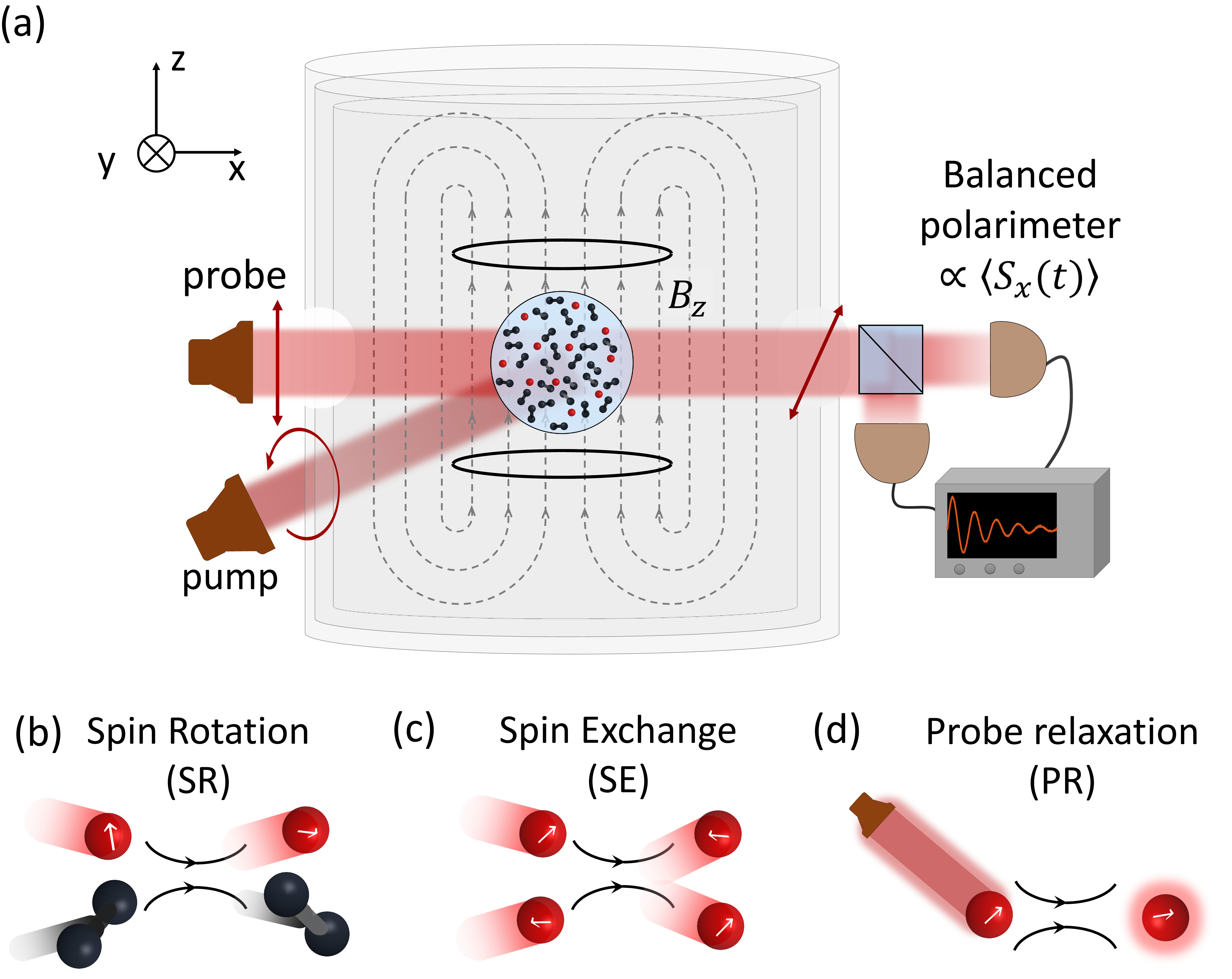}
\par\end{centering}
\centering{}\caption{\textbf{Relaxation mechanisms and experimental apparatus.} (a) Cesium spins (red spheres) are enclosed with a dense molecular N$_2$ gas (black spheres) inside a spherical glass cell. The cesium spins are initially polarized by a short pulse of optical pumping, and subsequently begin precessing around an applied magnetic field $B_z$. The spin evolution and coherence time is monitored using an optical probe beam, which measures the average spin signal $ S$ transverse to the magnetic field. (b-d) Processes which effect the electron spins' coherence time. (b) Randomization of the cesium's electron spin by the spin-rotation interaction during collisions with an N$_2$ molecule. (c) Exchange of electronic spins-between pairs of alkali metal atoms in collisions. (d) Absorption of near-resonance light which excites the cesium's electron and alters the spin state in the electronic ground-state after de-excitation. Unlike process (c) which instantaneously conserves the total spin of the colliding spins, processes (b) and (d) do not conserve but rather randomize the spin of the electron. \label{fig:system}}
\end{figure}

We experimentally study the decoherence of cesium vapor with N$_2$ gas using the apparatus in Fig.~\ref{fig:system}. 
The atoms are enclosed in a $1"$ diameter spherical glass cell, which contains $2.4$ amagat of $\textrm{N}_2$ gas and a small cesium-metal droplet whose temperature determines the vapor number-density. The temperature is controlled using a home-made oven to maintain a constant cesium vapor pressure, and the cell is magnetically shielded using several layers of magnetic shields. We characterize the relaxation mechanisms by monitoring the dynamics of the cesium spins absent the pumping beam, with the detailed experimental sequence and configuration described in \cite{SI}. We first optically pump the spins in the $xy$ plane using an auxiliary magnetic field and circularly polarized beam that is tuned on resonance with the $D_1$ optical transition. The latter, appears as a single spectral line with a linewidth of about $40$ %39.4
GHz owing to pressure broadening by the N$_2$ gas. Then, we turn off these fields and apply a magnetic field $B\hat{z}$ which leads to Larmor precession.  The average collective evolution of the alkali spins is monitored via measurement of the polarization rotation of a weak, far-detuned and linearly polarized optical probe beam using a 
balanced polarimetry detection signal $S$, which negligibly affects the evolution.

\begin{figure}[t]
\begin{centering}
\includegraphics[width=8.6cm]{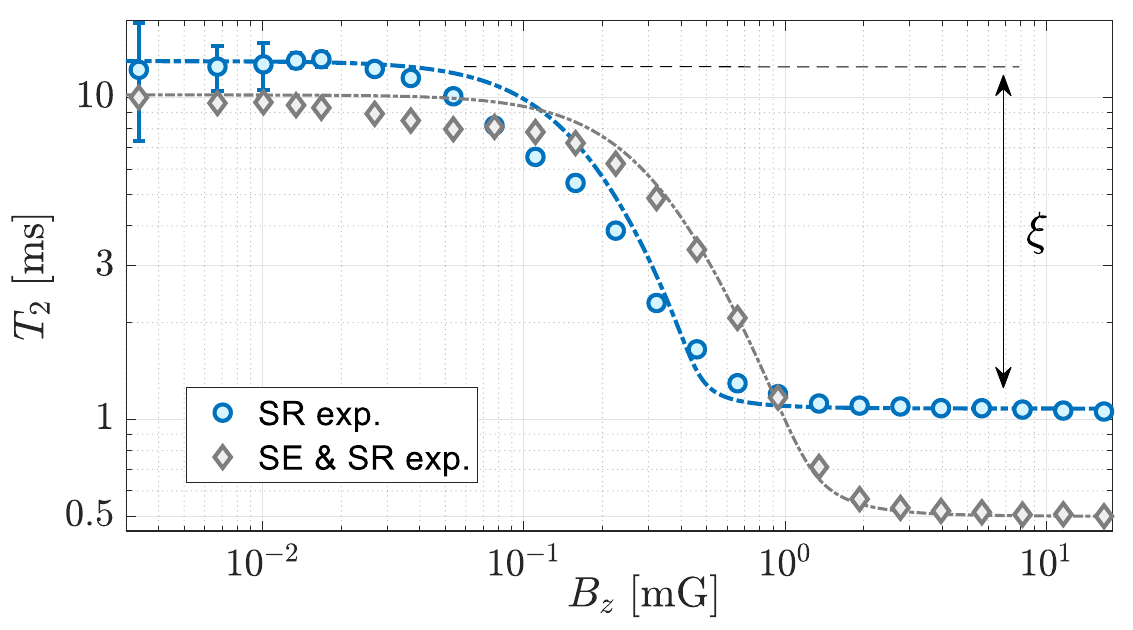}
\par\end{centering}
\centering{}\caption{\textbf{Suppression of decoherence by spin-rotation.} The measured coherence time $T_2$ at low magnetic fields is prolonged by the suppression factor $\xi$ with respect to high magnetic fields. We present two different configurations: lower vapor density (blue) where the suppression factor $\xi=12$ is predominantly associated with the spin-rotation relaxation, and higher vapor density (grey) where $\xi=20$ is associated with both the spin-rotation and spin-exchange which have comparable rates. The rates characterizing the two configurations are summarized in Table~\ref{tab:rates} under the columns SR and SE $\&$ SR, and the dashed lines correspond to the calculated relaxation rates as modeled in \cite{SI}. \label{fig:N2_destruction}}
\end{figure}

We measure the decoherence of the alkali metal spins as a function of the applied magnetic field $B\hat{z}$. For each value of the field, we fit the data to a simple model $ S(t) =A e^{-t/T_2}\sin{(\omega t+\phi)}$ using $A,\, \phi,\, \omega$ and $T_2$ as fitting parameters and $t$ to denote time of evolution. In Fig.~\ref{fig:N2_destruction} we present the measured coherence time $T_2$ as a function of the applied magnetic field $B\hat{z}$ (blue circles). We observe a $\xi=12$ fold suppression of the decoherence rate of the spins at low magnetic fields ($|B|\lesssim0.04$ mG) with respect to the decoherence at moderate magnetic fields ($|B|\gtrsim 0.8$ mG). 
These results show that the spins' decoherence rate can be suppressed by more than an order of magnitude at low magnetic fields.

To compare the contribution of the spin-rotation interaction to all other relaxation processes and confirm that it is the dominant relaxation mechanism, \BL{we perform independent estimations of the relevant relaxation rates} in our experiment as detailed in \cite{SI}, and summarized in Table \ref{tab:rates} under the "SR exp." column. Via measurement of the cesium number density $n_{\textrm{Cs}}=(3.3\pm0.3)\times10^{11}\,\textrm{cm}^{-3}$, we estimate the various spin-relaxation rates affecting the valence electron; The rate $R_{\textrm{sr}}^{(\textrm{N}_2)}$ by collisions of cesium and $\textrm{N}_2$, the spin-exchange rate $R_{\textrm{se}}$ due to collisions between pairs of alkali-metal atoms and the spin-destruction rate $R_{\textrm{pr}}$ by the weak probe beam. We also estimate and damping rate of the slowest diffusion mode $R_{\textrm{diff}}$ by diffusion and destruction at the enclosure walls. \BL{These independent estimations also agree well with the independently measured longitudinal spin lifetime $T_1=(15.0\pm0.5)$ ms}. Evidently, the spin-rotation rate $R_{\textrm{sr}}^{(\textrm{N}_2)}$ highlighted in blue is the dominant relaxation mechanism of the vapor, surpassing all other rates by about an order of magnitude. We therefore conclude that the suppressed decoherence observed in Fig.~\ref{fig:N2_destruction} is associated predominantly with the spin-rotation interaction.

We also study the suppression for different degrees of spin polarization and to different cesium number densities. The initial spin polarization ${P=T_1/(T_{\textrm{pump}}+T_1)}$ depends on the longitudinal optical-pumping time $T_{\textrm{pump}}$ which we control and independently measure \cite{SI}. We estimate an initial spin polarization of $P=0.15$ for the aforementioned experiment, which corresponds to the low-polarization limit. We repeat the experiment also for higher initial polarization $P=0.66$ and find similar behaviour and suppression factor $\xi$ \cite{SI}. 
We also characterize the suppression at an elevated number density $n_{\textrm{Cs}}=(7.5\pm0.5)\times10^{12}\,\textrm{cm}^{-3}$, in a configuration whose parameters appear under the "SE $\&$ SR exp." column in table~\ref{tab:rates}. Here the spin-exchange and spin-rotation rates are comparable, and their total measured decoherence, presented in in Fig.~\ref{fig:N2_destruction} (grey), is simultaneously suppressed at low magnetic fields. Frequent spin-exchange collisions also extend the range of usable magnetic fields for which the decoherence is maximally suppressed. 
These measurements demonstrate that the suppression of spin-rotation decoherence is robust for various degrees of spin polarization and alkali number densities, and that at low magnetic fields, the coherence time $T_2$ can be prolonged towards the spin-lifetime $T_1$.

\begin{figure}[t]
\begin{centering}
\includegraphics[width=8.6cm]{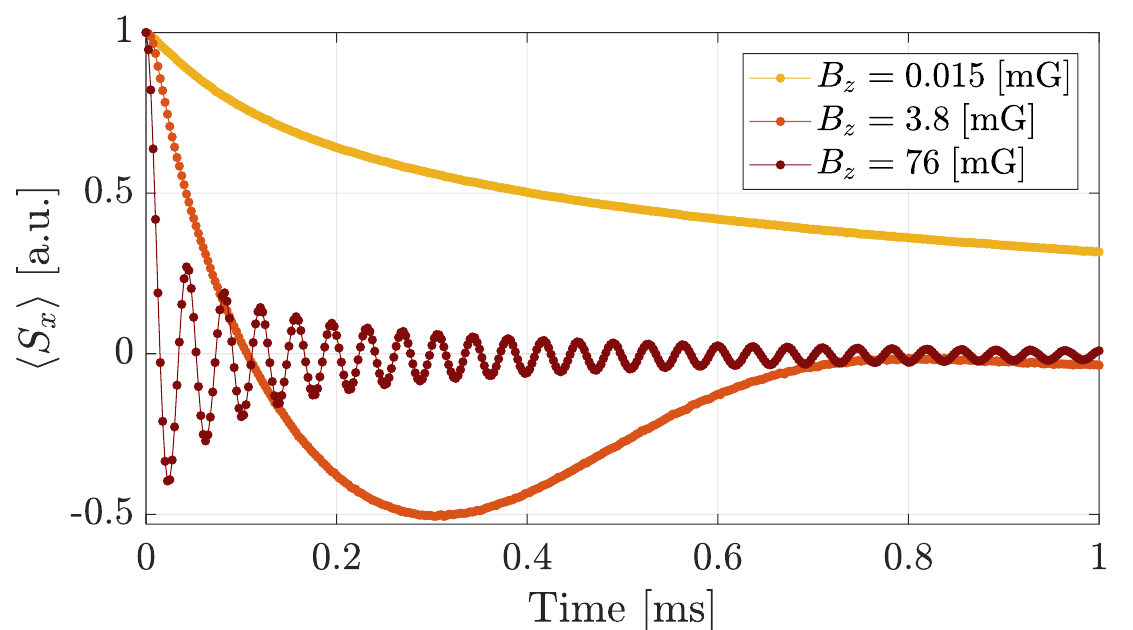}
\par\end{centering}
\centering{}\caption{\textbf{Suppression of decoherence by optical absorption.} The measured transverse spin relaxes rapidly at high magnetic fields, but the relaxation is suppressed at low magnetic fields. In this configuration the probe power is increased and tuned near resonance to render its relaxation dominant (see the experimental parameters in Table~\ref{tab:rates} under the PR exp.~coulmn). The measured decay is multi-exponential owing to the Gaussian shape of the probe beam.}  \label{fig:probe_destruction}
\end{figure}

We further study the spin-decoherence induced by absorption of near-resonant light. We repeat the experimental sequence for the parameters of "SR exp." but increase the optical power of the probe and tune its frequency near resonance with the optical D$_1$ transition to increase absorption. Consequently, the relaxation by the probe light $R_{\textrm{pr}}$ is considerably increased and dominates the spins relaxation. We also increase the pump power in the preparation stage to set the initial polarization $P=0.15$ despite the decreased lifetime $T_1=1.0\pm 0.06$ msec. In Fig.~\ref{fig:probe_destruction} we present the short-time evolution of the mean spin component $\langle S_x(t) \rangle$ at three different magnetic fields. Evidently, the decoherence at low magnetic field is suppressed with respect to higher magnetic fields, and approaches spin-lifetime $T_1$. Owing to the Gaussian intensity profile of the probe and the slow diffusion of the atoms, the decay is multi-exponential \cite{happer2010optically,appelt1998theory,shaham2020quantum} which precludes quantitative characterization by a single relaxation rate. Nonetheless, both the ratio of the linewidth of the signals in the frequency domain at high and low magnetic fields ($\xi =19 $) as well as the ratio between the $1/e$ time ($\xi=28$) indicate that the decoherence is suppressed by more than an order of magnitude. We repeat the experiment for variable optical powers and find similar results \cite{SI}.

\begin{table}[t]
\centering \includegraphics[width=8.6cm]{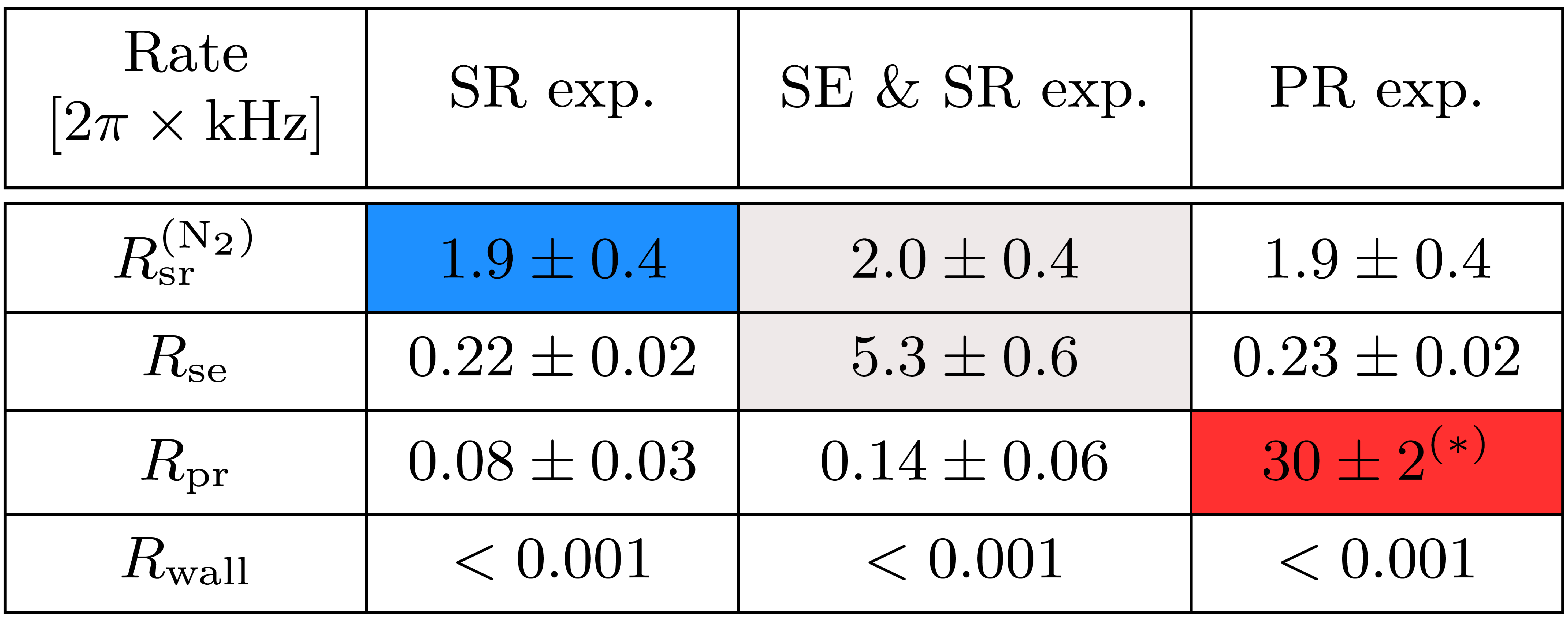}

 \caption{\textbf{Relaxation rates for the three experimental configurations.} $R_{\textrm{sr}}^{(\textrm{N}_2)}$ denotes the relaxation rate of the electron spin by collisions with $N_2$ molecules, $R_{\textrm{se}}$ denotes the spin exchange rate, $R_{\textrm{pr}}$ is the relaxation by of the electron spin due to light absorption of the probe beam. In the column PR this value was achieved from spin relaxation time at the first millisecond of the decay process under powerful probe beam. And $R_{\textrm{wall}}$ is the relaxation rate of the first diffusion mode by diffusion and interaction with the walls. The column SR corresponds to a configuration with low vapor density and low probe power, where the spin-rotation rate is dominant. The column SE $\&$ SR corresponds to a configuration with higher vapor density, where the rates of spin-rotation and spin-exchange are comparable. The coherence time of these two configuration is shown in Fig.~\ref{fig:N2_destruction} in blue and grey respectively. The column PR corresponds to low vapor density but high and near resonance optical probe beam, which renders the probe induced relaxation dominant. The evolution in the latter configuration is shown in Fig.~\ref{fig:probe_destruction}. Calculation of all rates is detailed in \cite{SI}}
 \label{tab:rates}
\end{table}

We first interpret these results using a simple model \cite{happer1972optical,appelt1998theory,katz2015coherent,xiao2021atomic}. In alkali-metal atoms, the strong hyperfine interaction couples the electron spin $S$ and the nuclear spin $I$, rendering the total spin $F=I+S$ and its projection along the magnetic field $M$ to good quantum numbers. We describe the evolution of $\langle F_{+}\rangle=\langle F_{x}\rangle+i\langle F_{y}\rangle$ and $\langle S_{+}\rangle=\langle S_{x}\rangle+i\langle S_{y}\rangle$, associated with the total and electron transverse spin of an alkali-metal atoms respectively. Subject to magnetic field $B_z\hat{z}$ and assuming time-averaged relaxation of the electron spin at a rate $R_{\textrm{sd}}$ the evolution is given by
\begin{align}\partial_{t}\langle F_{+}\rangle &=-\left(ig_{e}B+R_{\textrm{sd}}\right)\left\langle S_{+}\right\rangle, \label{eq:dF_dt}\\\partial_{t}\left\langle S_{+}\right\rangle &=-i\frac{g_{e}B}{2q}\left\langle F_{+}\right\rangle -R_{\textrm{sd}}(\left\langle S_{+}\right\rangle-\tfrac{1}{q}\left\langle I_{+}\right\rangle).\label{eq:dS_dt}\end{align}
Eq.~(\ref{eq:dF_dt}) describes the precession of the total spin induced by the Larmor precession of the electron spin where $g_e$ is the electron gyromagnetic-ratio, as well as the decay due to damping of the electron spin at a rate $R_{\textrm{sd}}$. The first term in Eq.~(\ref{eq:dS_dt}) describes the back-action of the total spin on the electron which slows down its motion, the second term describes its damping, and the last term describes the flywheel repolarization  by a fraction $q=(2I+1)^{2}/2$, known as the slowing-down factor, via the transverse nuclear spin $\langle I_{+}\rangle=\langle F_{+}\rangle-\langle S_{+}\rangle$.

Equations.~(\ref{eq:dF_dt}-\ref{eq:dS_dt}) form a coupled set of two linear differential equations whose dynamics is characterized by the evolution of two modes 
corresponding to the eigenvalues \begin{equation}\label{eq:3}
\lambda_{\pm}=-\frac{R_{\textrm{sd}}}{2q}\left(1+q\pm\sqrt{\left(1-q\right)^{2}-2qx\left(i+x\right)}\right),
\end{equation}
with $x=g_eB/R_{\textrm{sd}}$. These eigenvalues determine the precession rates $\omega=\textrm{Im}(\lambda_\pm)$ and decoherence rates $T_2^{-1}=-\textrm{Re}(\lambda_\pm)$ of the modes as a function of the magnetic field. At high magnetic fields ($x\gg1$) we find two precession rates $\omega=\pm g_eB/(2I+1)$, and two rapid decoherence rates, $T_2^{-1}=R_{\textrm{sd}}(q\pm\sqrt{2q}+1)/(2q)$. These correspond to two modes associated with the transverse magnetic dipole moment in the two hyperfine manifolds; $\lambda_+\,(\lambda_-)$ is associated with the upper (lower) hyperfine manifold, where the electron and nuclear spins are aligned (anti-aligned), consistent with \cite{bouchiat1963relaxation,happer1972optical}. At low magnetic fields ($x\ll1$), the precession frequencies of the two modes slow to $\omega=\pm g_eB/(2q-2)$, and the decoherence rates become $R_{\textrm{sd}}$ and $R_{\textrm{sd}}/q$, with the latter exhibiting suppressed relaxation.

The two-mode decomposition leads to an oscillatory, bi-exponential decay of the spin signal. The degree of actual observed suppression at low magnetic fields is associated with the dominance of the slowly decaying mode, which depends on the initial spin state of the ensemble and on the measured observable. For our configuration, the experimental signal is mostly sensitive to the slowly decaying mode, explaining the high-observed suppression and in agreement with our theoretical analysis \cite{SI}. In configurations where the dynamics are fully captured by the slow mode, one would expect $\xi\approx q/2\approx16$ for cesium.

It is intriguing to compare the transverse relaxation $T_2^{-1}$ we measure with the longitudinal one $T_1^{-1}$. While $T_2^{-1}$ strongly depends on the magnetic field [Eq.~(\ref{eq:3})], $T_1^{-1}$ is magnetic field-independent until extreme fields \cite{walter2002magnetic}. However, in the $B=0$ limit, the two characteristic times align due to the isotropy of the relaxation process and the absence of a quantization axis, rendering them degenerate $T_1=T_2$. In this limit, the results agree with the theoretical expressions for the longitudinal spin lifetimes of the fundamental mode \cite{appelt1998theory, walter2002magnetic}. 

We extend the previous analysis and include the effect of spin-exchange assuming small spin polarization $P\ll1$, as well as the small contribution of diffusion-damping  \cite{SI}. The modeled response is shown in Fig.~\ref{fig:N2_destruction} (dot-dashed lines), and is in good agreement with the measured data. \BL{This agreement further supports the accuracy of our rate identifications.} At higher spin-exchange rates, the range of magnetic field values over which relaxation is suppressed expands. Spin-exchange increases decoherence at high magnetic fields, alters the longitudinal lifetime \cite{appelt1998theory}, and affects the low-field decoherence rate. In the SERF regime, sensors benefit from suppressed decoherence due to both spin-exchange and spin-destruction processes. For low polarization and $R_{\textrm{se}}\gg R_{\textrm{sd}}$, we estimate the fundamental decoherence rate of cesium-atom SERF magnetometers to be approximately $45\%$ higher than in the $R_{\textrm{se}}\ll R_{\textrm{sd}}$ regime studied here \cite{SI}.  

The simple model describes relaxation arising solely from the electron spin $S$, akin to that caused by white magnetic field noise coupled exclusively to it \cite{happer2010optically,happer1972optical}. It assumes the noise correlation time is much shorter than the hyperfine interaction timescale, effectively "freezing" the nuclear spin during the instantaneous perturbation. For spin-rotation interactions, collisions last a few picoseconds \cite{happer1987optical,happer2010optically}, with mean intervals of tens of picoseconds under the gas pressure used here: shorter than and comparable to the hyperfine timescale, respectively. In the pressure-broadened optical transition, probe-induced relaxation involves photon absorption and non-radiative de-excitation via $\textrm{N}_2$ collisions, with correlation times of a few picoseconds. While such $S$-damping explains our results well, deviations could offer insights into relaxation mechanisms at low magnetic fields.

In conclusion, we have experimentally demonstrated the magnetic field dependence of $T_2$-type decoherence for processes that do not preserve spin, such as spin-rotation interaction and light absorption, exhibiting more than an order of magnitude variation. This work extends the use of magnetic fields as an effective tool to suppress decoherence originating from processes that predominantly affect the electron spin of alkali-metal atoms.

The dependence of relaxation rates on the magnetic field could find applications in various contexts. Firstly, it could play a role in the design of atomic-based sensors, such as the optimal pressure of miniature magnetometers we study in \cite{SI}. Secondly, the variation in magnetic field strength holds potential in quantum information applications. For instance, in optical quantum memories, which typically involve mapping the photonic state onto and from the spin ensemble in the presence of a magnetic field \cite{noble-gas-PRA,Polzik-coherent-state-memory,Polzik-squeezed-states-memory}, reducing the magnetic field could potentially prolong the storage time if relaxation is governed by collisions that relax the electronic spin. In other configurations, such as those explored in \cite{moller2017quantum,thomas2021entanglement,khalili2018overcoming,julsgaard2001experimental,katz2020long}, the Faraday interaction with the probe beam enables the generation of entanglement, albeit often limited by the absorption of off-resonant light. The relaxation induced by the latter could potentially be suppressed at low magnetic fields. Lastly, our findings may have broader applications, such as in cold atomic gases trapped using off-resonant optical traps \cite{cline1994spin}. In these setups, photon scattering by absorption serves as a fundamental relaxation mechanism, primarily relaxing the electron spin due to the large detuning relative to the spectral excited state lines. Here, operating at low magnetic fields could potentially slow down the relaxation of the total spin.

It is also interesting to consider the potential suppression of decoherence by other relaxation processes. The relaxation mechanisms we considered here coupled predominantly through the spin of the electron, owing to the short correlation time of the interaction. Other relaxation processes also have sizeable coupling to the electron spins. These processes include relaxation by spin-axis in collisions of alkali-metal pairs \cite{erickson2000spin,kadlecek2001spin}, relaxation by the walls of anti-relaxation coated cells \cite{balabas2010polarized}, and relaxation by very short-lived van-der-Waals molecules \cite{nelson2001rb}. It is therefore plausible that the decoherence rate by these processes would also be suppressed at lower magnetic fields. 

\bibliography{Refs}

\end{document}